\documentclass[12pt]{article}

\usepackage{epsfig}
\usepackage{amsbsy}
\usepackage{amsmath}

\begin{document}
\title{Bose-Einstein correlations: a study of an invariance group \thanks{Supported in part by the KBN grant 2P03B
093 22}}
\author{A.Bialas and K.Zalewski
\\ M.Smoluchowski Institute of Physics
\\ Jagellonian University, Cracow
\\ and\\ Institute of Nuclear Physics, Cracow}
\maketitle

\begin{abstract}
A group of transformations changing the phases of the elements of the
single-particle density matrix, but leaving unchanged the predictions for
identical particles concerning the momentum distributions, momentum
correlations etc., is identified. Its implications for the determinations of
the interaction regions from studies of Bose-Einstein correlations are
discussed.
\end{abstract}
\noindent PACS numbers 25.75.Gz, 13.65.+i \\Bose-Einstein correlations,
interaction region determination.
\section{Introduction}
Much work is being done on Bose-Einstein and Fermi-Dirac correlations
in multiparticle production processes. The purpose in most cases is to
determine the features of the interaction region i.e. of the region where
hadrons are produced. Let us suppose that function $\tilde{\rho}(\textbf{x})$
is the density of some kind of hadrons, say of negative pions, just after they
have been produced. Knowing this function one would know the geometry of the
interaction region. According to textbook formulae $\tilde{\rho}(\textbf{x})$
is proportional to the diagonal elements of the density matrix\footnote{It is
simplest to use the interaction picture. Then the density matrix after
freeze-out is time-independent} in the coordinate representation
$\rho_c(\textbf{x};\textbf{x})$. In multiple particle production processes,
however, these matrix elements cannot be directly measured -- there is no ruler
to measure the interaction region. One measures momenta and uses the relation
between the density matrices in the coordinate and the momentum representations

\begin{equation}\label{}
  \rho_c(\textbf{x};\textbf{x}) = \int \frac{d^3p_1d^3p_2}{(2\pi)^3}\rho(\textbf{p}_1;
  \textbf{p}_2)e^{i\textbf{q}\textbf{x}},
\end{equation}
where

\begin{equation}\label{}
\textbf{q} = \textbf{p}_1 - \textbf{p}_2
\end{equation}

Thus, in order to determine the shape of the interaction region it is
necessary to know both the diagonal and the out-of-diagonal elements of
the density matrix in the momentum representation. The diagonal elements
are easily obtained, because they are proportional to the density of the
single particle momentum distribution $\Omega(\textbf{p})$. The only
thing one can do to get information about the out-of-diagonal elements
is to measure the two-body, three-body etc. momentum distributions. In
general it is not possible to express these distributions in terms of
the single particle density matrix. Making, however, the usual
assumption that the $k$-particle density matrix can be approximated by
the symmetrized product of single particle density matrices  (see, e.g.,
\cite{KAR,BIK,CSO}), one can determine from the momentum distributions
of up to $k$ particles the functions

\begin{equation}\label{measur}
  \Re\left[\rho(\textbf{p}_1;\textbf{p}_2)
  \rho(\textbf{p}_2;\textbf{p}_3)\ldots\rho(\textbf{p}_k;\textbf{p}_1)\right],
\end{equation}
where $\Re$ stand for real part of. Thus from single particle momentum
distributions one gets the diagonal elements $\rho(\textbf{p}_1;\textbf{p}_1)$
as already mentioned. Adding the information about two-body momentum
distribution one finds further
$\rho(\textbf{p}_1;\textbf{p}_2)\rho(\textbf{p}_2;\textbf{p}_1) =
|\rho(\textbf{p}_1;\textbf{p}_2)|^2$ -- the absolute values of the
out-of-diagonal matrix elements, but not their phases. Measuring more-particle
momentum distributions one finds some further information abut the phases, but
not all of it. It turns out that it is possible to include in the
single particle density matrix elements an additional phase factor without
changing the predicted momentum distributions.  The choice of this factor,
which is unconstrained by the data, has a dramatic effect on the size and shape
of the interaction region which one deduces from the momentum measurements. It
is the purpose of the present paper to investigate this effect in a systematic
way.

In the next section the phase ambiguity is explicitly formulated. Its
effect on the size and shape of the interaction region is discussed in Section
3. Two specific examples are presented in Sections 4 and 5. Our conclusions are
summarized in the last section.

\section{Group of transformations leaving the momentum distributions invariant}
It is easily checked that making the replacement

\begin{equation}\label{modrho}
  \rho(\textbf{p}_1;\textbf{p}_2) \rightarrow \rho_\phi(\textbf{p}_1;\textbf{p}_2) =
  e^{i[\phi(\textbf{p}_1)-\phi(\textbf{p}_2)]}\rho(\textbf{p}_1;\textbf{p}_2),
\end{equation}
where  $\phi(\textbf{p})$ is
an arbitrary, real-valued\footnote{When $\phi$
is real-valued,  matrix $\rho_\phi$ is hermitian and has trace one,
thus it can be interpreted as a single particle density matrix.},
 function, one leaves all the observables (\ref{measur})
unchanged. Thus the measured momentum distributions are invariant with respect
to the   transformation given by (\ref{modrho}). In
other words, the measured momentum distributions yield no argument for or
against any choice of $\phi(\textbf{p})$.

Before analyzing the effect of this phase ambiguity on the determination of the
interaction region let us recall the standard procedure.

 In order to obtain information on the interaction region
 from the Bose-Einstein correlations in multiple particle production one
 usually applies the formula (cf. e.g. the review \cite{WIH})

\begin{equation}\label{emifun}
  \rho(\textbf{p}_1;\textbf{p}_2) = \int d^4X S(K,X)e^{iqX},
\end{equation}
where

\begin{equation}\label{}
  K = \frac{1}{2}(p_1 + p_2),\qquad X = \frac{1}{2}(x_1 + x_2)
\end{equation}
and $S(K,X)$ is known as the emission function. It follows from the hermiticity
of the density matrix that the emission function is real-valued. It is
constructed by analogy with the Wigner function and is interpreted as the
position and momentum distribution of the produced hadrons \cite{WIH},
\cite{PRA}. As is well known, this interpretation cannot be exact\footnote{
Taken literally, it contradicts the uncertainty principle. The solution is also
well known: one has to perform an adequate "smearing" of $S(K,X)$ bringing it
into agreement with the requirements of quantum mechanics.}. We will accept it
here,  however, without further discussion.

Another well-known difficulty (see e.g. \cite{ZAL}) in the studies of the
interaction region is that the Fourier transform (\ref{emifun}) cannot be
inverted. The reason is that the density matrix on the left-hand side is known
only for the on-mass-shell values of the four-momenta $p_1$,$p_2$. In general,
there is an infinite variety of different emission functions which correspond
to a given density matrix. In order to avoid getting involved with this
difficulty  we limit our discussion to models, where the emission function can
be unambiguously expressed by the Wigner function. Two such cases have been
much discussed. When all the identical particles are produced simultaneously at
some time $t = 0$, one can choose

\begin{equation}\label{}
  S(K,X) = \delta(X_0)W(K,\textbf{X}),
\end{equation}
where $W$ is the Wigner function corresponding to the time-independent
(interaction representation) density matrix in the momentum representation.
When the distribution of longitudinal momenta is weakly correlated with the
distribution of transverse momenta, one can consider the transverse emission
function $S_T(\textbf{K}_T,\textbf{X}_T)$ as the Wigner function corresponding
to the transverse factor $\rho_T(\textbf{p}_{T1};\textbf{p}_{T2})$ in the
density matrix \footnote{For an example see Section 5.}. We will thus use the
relation

\begin{equation}\label{wigrho}
  W_\phi(\textbf{K},\textbf{X}) =
  \int \tilde{d\textbf{q}}\rho_\phi(\textbf{K},\textbf{q})
e^{i\textbf{qX}},
\end{equation}
where $\rho_\phi(\textbf{K},\textbf{q}) \equiv
\rho_\phi(\textbf{p}_1;\textbf{p}_2)$, $\tilde{d{\bf q}}\equiv
d\textbf{q}/(2\pi)^l$ and $d{\bf q} = d^lq$ with $l$ being the number of
dimensions considered ( the vectors are two- or three-dimensional depending on
the model). The Wigner function $W_\phi(\textbf{K},\textbf{X})$, given by
(\ref{wigrho}), will be used to define the interaction region.

\section{The phase and the moments of the $X$ distribution}

In order to study  the effects of the full group of transformations
(\ref{modrho}), let us first discuss  the moments of the distribution of
$\textbf{X}$. The $K$-dependent averages, for a given choice of the phase
$\phi$, are {\bf denoted} by

\begin{equation}\label{averka}
  \langle g(\textbf{X}) \rangle_\phi (\textbf{K}) = \frac{\int d\textbf{X} W_\phi(\textbf{K},\textbf{X})g(\textbf{X})}{\int d\textbf{X}
  W_\phi(\textbf{K},\textbf{X})}
\end{equation}
and the full averages by

\begin{equation}\label{avrtot}
  \langle\langle g \rangle\rangle_\phi = \int d\textbf{K} d\textbf{X}
  \textbf{W}_\phi(\textbf{K},\textbf{X})g(\textbf{X}).
\end{equation}
Since in the latter average the Wigner function has been integrated over
$\textbf{K}$, there is no problem with the uncertainty principle\footnote{
There may be, however, problems with the interpretation of the averages
$\langle g(\textbf{X}) \rangle(\textbf{K})$.}. It is convenient to introduce
the notation

\begin{eqnarray}\label{}
\langle \textbf{X} \rangle_{\phi=0}(\textbf{K}) &= \textbf{r}_0(\textbf{K});
\qquad \langle \textbf{X}^2 \rangle_{\phi=0} &= R_0^2({\bf K});\\
\langle\langle \textbf{X} \rangle\rangle_{\phi=0} &= \textbf{r}_0;\qquad
\langle\langle \textbf{X}^2 \rangle\rangle_{\phi=0} &= R_0^2.
\end{eqnarray}

Using the definition (\ref{wigrho}) and the identity (one dimension,
$n=0,1,\ldots$)

\begin{equation}\label{}
  \int dx x^n e^{iqx} = (-i)^n\frac{\partial^n}{\partial q^n}\int dx e^{iqx} =
  2\pi (-i)^n \frac{\partial^n}{\partial q^n} \delta (q).
\end{equation}
one finds that the denominator on the right-hand side of (\ref{averka}) equals
$\rho_\phi(\textbf{K},0)$, and

\begin{equation}\label{14}
\langle X_j^n \rangle_{\phi}(\textbf{K}) =
\frac{i^n}{\rho_\phi(\textbf{K},\textbf{0})}\left(\frac{\partial^n}{\partial
q_j^n}\rho_\phi(\textbf{K},\textbf{q})\right)_{\textbf{q}=0},
\end{equation}
which is easily generalized to

\begin{equation}\label{}
\langle \prod_j X_j^{n_j} \rangle_{\phi}(\textbf{K}) =
\frac{i^n}{\rho_\phi(\textbf{K},\textbf{0})}\left(
\frac{\partial^n}{\prod_j\partial q_j^{n_j}} \rho_\phi(\textbf{K},\textbf{q})
\right)_{\textbf{q}=0},
\end{equation}
where $n = \sum_j n_j$.

When evaluating these averages we will need derivatives of the phase factor

\begin{equation}\label{derq}
\phi(\textbf{p}_1)-\phi(\textbf{p}_2) = \phi(\textbf{K} + \frac{\textbf{q}}{2})
  - \phi(\textbf{K} - \frac{\textbf{q}}{2}) = 2\sinh(\frac{\textbf{q}\cdot
  \nabla}{2})\phi(\textbf{K}),
\end{equation}

These formulae allow to calculate all moments $\langle\langle X_j^n
\rangle\rangle$ in terms of  (the integrals of) the derivatives of the single
particle density matrix. This gives the full information about the size
and shape of the interaction region.

\section {Position-momentum correlations and the size of the interaction
region}

We will now discuss in detail two parameters characterizing the interaction
region: (i) the average position $\langle\langle \textbf{X} \rangle\rangle$ and
(ii) the averaged square of the size $\langle\langle \textbf{X}^2
\rangle\rangle - \langle\langle \textbf{X} \rangle\rangle^2$.

 Using the formulae\footnote{They are special cases of (\ref{derq}).}
\begin{equation}\label{}
  \left[\nabla\left(\phi(\textbf{p}_1)-\phi(\textbf{p}_2)\right)\right]_
{\textbf{q} = 0} =
  \nabla \phi(\textbf{K});\qquad
  \left[\nabla^2\left(\phi(\textbf{p}_1)-\phi(\textbf{p}_2)\right)\right]_
{\textbf{q} = 0} =0
\end{equation}

 one finds

\begin{eqnarray}\label{expalf}
  \langle \textbf{X} \rangle_\phi(\textbf{K}) &=& \textbf{r}_0(\textbf{K}) -
\nabla \phi(  \textbf{K}),\\
  \langle \textbf{X}^2 \rangle_\phi(\textbf{K}) &=& R_0^2(\textbf{K}) - 2
  \textbf{r}_0(\textbf{K})\cdot\nabla \phi(\textbf{K}) +
  \left(\nabla \phi(\textbf{K})\right)^2
\end{eqnarray}

and

\begin{eqnarray}\label{}
  \langle\langle \textbf{X} \rangle\rangle_\phi &=& \textbf{r}_0
 - \langle\nabla \phi(  \textbf{K})\rangle\\
  \langle\langle \textbf{X}^2 \rangle\rangle_\phi &=& R_0^2 -
  2\langle
  \textbf{r}_0(\textbf{K})\cdot\nabla \phi(\textbf{K})\rangle+
  \langle\left(\nabla \phi(\textbf{K})\right)^2\rangle,
\end{eqnarray}
where the notation

\begin{equation}\label{}
  \langle g(\textbf{K}) \rangle =\int
  d\textbf{K}\rho_\phi(\textbf{K},0)g(\textbf{K})
\end{equation}
has been used.

Function $\langle \textbf{X} \rangle_\phi(\textbf{K})$ can be
interpreted as a measure of the position-momentum correlations.
 According to formula (\ref{modrho}) the phase  is
fixed, when the function $\phi$ is given.  On the other hand, as
seen from formula (\ref{expalf}),  $\phi({\bf p})$ can be also fixed,
up to an uninteresting constant, by specifying the position momentum
correlation $\langle \textbf{X} \rangle (\textbf{K})$.

 The variance of $\textbf{X}$ at fixed $\textbf{K}$ is of particular
interest because it gives the HBT radius, {\it as determined in
the usual way from the two-particle distribution in the Gaussian
approximation}.  This can be seen as follows. First, we observe
that it  is invariant, i.e. does not depend on the choice of
$\phi({\bf p})$ :
\begin{equation}\label{vargik}
 \langle \textbf{X}^2 \rangle_{\phi}(\textbf{K}) -
\left(\langle \textbf{X}
 \rangle_{\phi}(\textbf{K})\right)^2=R_0^2(\textbf{K}) - r_0^2(\textbf{K}).
\end{equation}
It can be therefore calculated independently of the choice of
$\phi({\bf p})$. On the other hand, in the Gaussian approximation
\begin{equation}
\rho({\bf K},{\bf q})=\rho({\bf K},{\bf q}=0)e^{-{\bf q}^2R_{HBT}^2({\bf
K})/2}
\end{equation}
where $R_{HBT}^2({\bf K})$ is defined by this relation.
Formula (\ref{14}) implies for this approximation $\langle{\bf X}\rangle({\bf
K})=0$, $\langle X_i^2\rangle({\bf K})=R_{HBT}^2({\bf K})$. According to
(\ref{vargik}) we thus have
\begin{equation}
d R_{HBT}^2({\bf K}) =  \langle \textbf{X}^2 \rangle(\textbf{K}) -
\left(\langle \textbf{X}
 \rangle(\textbf{K})\right)^2=R_0^2(\textbf{K}) - r_0^2(\textbf{K}).
\end{equation}
where $d=2,3$ is the number of dimensions.

Note, however, that the variance of ${\bf X}$ at fixed ${\bf K}$
(and thus also the measured HBT radius) is not  the {\it physical}
size of the interaction region at fixed ${\bf K}$. It is only an
auxiliary concept. Indeed, according to Heisenberg's uncertainty
principle fixing ${\bf K}$ implies that the true variance of ${\bf
X}$ becomes infinite (cf. footnote 3).

The true size of the interaction region, as derived from the Wigner
function, is  instead determined  by the full average, given by

\begin{eqnarray}\label{varpos}
R^2_{true}\equiv \langle\langle \textbf{X}^2 \rangle\rangle_\phi - \langle\langle \textbf{X}
  \rangle\rangle_\phi^2 =
 R_0^2 - \textbf{r}_0^2 +\nonumber \\+ \langle [\nabla \phi({\bf K})]^2\rangle
  - \langle (\nabla \phi({\bf K}))\rangle^2
- 2(\langle \textbf{r}_0(\textbf{K})\nabla \phi({\bf K})\rangle_\textbf{K} -
  \textbf{r}_0\langle \nabla \phi({\bf K}) \rangle)
\end{eqnarray}
One sees that this quantity explicitly depends on $\phi({\bf p})$ and
so does the interpretation of the HBT measurements.

The three pairs of terms on the right-hand side are easily  identified as: the
variance at fixed $\phi = 0$, the variance of $\nabla \phi$, and twice the
unnormalized correlation coefficient for $r_0(\textbf{K})$ and $\nabla  \phi$. The
variance of $ \phi$ increases the observed variance. The correlation may
have either sign. In particular, it may be negative and so large in absolute
value that it reduces the observed variance below the variance at $ \phi=0$. This
is not surprising. Suppose that the variance for some density matrix $\rho_0$ is
smaller than for $\rho_\phi$. Then one could start with $\rho_\phi$ and go
over to the original $\rho_0$ by introducing a suitable phase. This change of
phase reduces the variance.

 Equating to zero the (variational) derivative of the right-hand side of
equation (\ref{varpos}) with respect to $\nabla  \phi$ one finds the
condition for the variance of $\textbf{X}$ (i.e. the size of the
interaction region) to be minimal. It turns out that the minimum is
achieved when $ \phi$ is chosen such that the average of $\textbf{X}$ at
fixed ${\bf K}$ does not depend on ${\bf K}$:

\begin{equation}\label{xk}
  \langle \textbf{X} \rangle_\phi(\textbf{K}) =
\textbf{r}_0(\textbf{K}) - \nabla  \phi(
  \textbf{K})={\bf a},
\end{equation}
where $\textbf{a}$ is a constant vector.
In other words, the minimal size is obtained when the
phase is chosen such that there is no position-momentum correlation. It is clear from
(\ref{xk}) that one can always find $ \phi$ which satisfies this condition.

Substituting (\ref{xk}) into (\ref{expalf}) one finds that the minimal
size of the system (for the given momentum distribution) is

\begin{eqnarray}\label{}
\left[\langle\langle \textbf{X}^2 \rangle\rangle_\phi - \langle\langle \textbf{X}
  \rangle\rangle_\phi^2\right]_{min} &=& R_0^2 - \langle \textbf{r}_0^2(\textbf{K})
 \rangle =<R_{HBT}^2({\bf K})>  .
\end{eqnarray}
For a generic $\phi(\textbf{p})$ the interaction regions at various
$\textbf{K}$'s are shifted with respect to each other. This increases the
overall, integrated over $\textbf{K}$, interaction region. Condition $\langle
\textbf{X} \rangle_{\phi}(\textbf{K}) = \textbf{a}$ puts all these partial integration
regions, as well as possible, on top of each other. This minimizes the overall,
calculated interaction region. The minimum is just the average over
$\textbf{K}$ of expression (\ref{vargik}).  We thus conclude that the
 HBT radius, averaged over ${\bf K}$, corresponds to the minimal
possible size of the interaction region.

In order to find the moments up to order $2m$ it is enough to know the
odd-order derivative of $ \phi$ up to $\nabla^{2m-1} \phi$. Thus $\langle \textbf{X}
\rangle$ and $\langle X_iX_j \rangle $ can be calculated when $\nabla  \phi$ is
known. Knowing moreover $\nabla^3 \phi$ one can calculate all the moments up to
fourth order. In general it is not possible to calculate the Wigner function
$W_\phi$ in closed form. We will discuss, however, two instructive special
cases.

\section{Example I}
 We shall consider only the transverse variables. Choosing
\begin{equation}\label{linear}
   \phi(\textbf{p}) = a +\sum_i b_i p_i +\sum_i c_i p_i^2,
\end{equation}
with $a, b_i, c_i$ constant $(i=x,y)$, one obtains
\begin{equation}\label{}
\phi({\bf p}_1)-\phi({\bf p}_2) =\sum_i q_i(b_{i}+ 2 c_{i}K_i)\equiv
\sum_iq_iV_i.
\end{equation}
According to formula (\ref{expalf}) this is the most general case where
the phase $\phi({\bf p}_1)-\phi({\bf p}_2)$ is linear in $q$.
 In this case, from (\ref{emifun}), we can write

\begin{equation}\label{}
  \rho_\phi(\textbf{p}_1;\textbf{p}_2) = \int d^2X S({\bf K,X})
e^{i{\bf q}({\bf X}+{\bf V})}.
\end{equation}
Changing the integration variables ${\bf X}$ into ${\bf X}+ {\bf V}$ one finds the
emission function corresponding to the density matrix $\rho_\phi$

\begin{equation}\label{emishi}
  S_\phi({\bf K,X}) = S\left({\bf K,X-V})\right)
\end{equation}
Thus, the interaction region is shifted in space-time by ${\bf V}$:

\begin{equation}\label{}
  X_i \rightarrow X_i - b_{i} - 2 c_{i}K_i.
\end{equation}

This formula shows that introducing the phase (\ref{linear}) may have two
effects:

\begin{itemize}
  \item A rigid shift of the interaction region.
  \item A correlation between the momentum and the space-time position of the
emission point of the particle.
\end{itemize}
The first point is well-known and not really disturbing. It has been
discussed by many authors. The second has some relation to the notion of
the homogeneity region \cite{BOW,SIN}, but to the best of our
knowledge its relation to the phase of the density matrix has not been
pointed out.

In order to show that these considerations are not purely academic we will now
consider a specific model \cite{BIZ,BKP1,BKP2} where the
emission function is derived from a physical picture. The main assumptions, at
the classical level, are

\begin{equation}\label{hubble}
  K^\mu = \lambda X^\mu\qquad \mbox{for} \qquad \mu = 0,1,2,3,
\end{equation}

\begin{equation}\label{deftau}
X_0^2 - X_\|^2 = \tau^2,
\end{equation} 
where $\tau$ is a constant. Assumptions (\ref{hubble}) and
(\ref{deftau}) for the temporal and longitudinal components were
introduced by Bjorken \cite{BJO} and Gottfried \cite{GOT} in their boost
invariant description of the multiple particle production processes. The
full assumption (\ref{hubble}) was introduced by Cs\"org\"o and
Zim\'anyi \cite{CSZ} who used it to explain why in $e^+e^-$
annihilations the correlation function depends on $q^2$ only. The
motivation in \cite{BIZ} was to explain why the estimated radii of the
interaction region decrease rapidly with increasing particle mass. Let
us add the definition of the transverse mass

\begin{equation}\label{}
  M_T^2 = K_0^2 - K_\|^2.
\end{equation}
Note that, while for a particle of given mass $m_T^2 = m^2 + p_T^2$ depends on
the transverse momenta, here $K^2$ is not fixed and thus $M_T^2$ depends on the
longitudinal and temporal components of momentum. Assumptions (\ref{hubble})
and (\ref{deftau}) imply

\begin{equation}\label{}
  \lambda = \frac{M_T}{\tau}.
\end{equation}
It is easy to see qualitatively, why in this model the measured radius of the
interaction region is a decreasing function of the particle mass. Let us make
the crude assumption that the momentum distributions are the same for all the
particles. Then the distribution of $M_T X$ is universal and since $M_T$ is
bigger for heavy particles, the corresponding distribution in $X$ is narrower.

 Let us consider the
transverse part of the emission function proposed in \cite{BKP1}:

\begin{equation}\label{}
  S_T = \exp\left[-\frac{\textbf{X}_T^2}{2r_T^2} - \frac{\left(\textbf{K}_T -
  \frac{M_T}{\tau}\textbf{X}_T\right)^2}{2\delta_T^2}\right].
\end{equation}
It can be interpreted as a Wigner function\footnote{At fixed $M_T$} and thus
our previous discussion is applicable to it. This Wigner function can be
rewritten in the form

\begin{equation}\label{}
  S_T = \exp\left[-\frac{\phi_T^2}{2R_D^2} - \frac{\left(X_T -
  \phi_T\right)^2}{2R_\phi^2}\right],
\end{equation}
where

\begin{eqnarray}\label{}
  R_\phi &=& \frac{r_T}{\sqrt{1 + \mu^2}}\\
  \phi_T &=& r_T\frac{\mu}{1 + \mu^2}\frac{\textbf{K}_T}{\delta_T}\\
  R_D &=& \mu R_\phi
\end{eqnarray}
with

\begin{equation}\label{}
  \mu = \frac{r_T M_T}{\tau \delta_T}.
\end{equation}
$S_T$ and $\phi_T$ are of the form discussed in the preceding
section\footnote{Because of the presence of $M_T$ this is an approximation,
valid when only two-body symmetrization is used and/or when the variance of
$M_T$ can be neglected.}. We thus conclude that the \textit{measured} HBT
radius is $R_\phi$, whereas the \textit{true} size of the interaction region,
according to (\ref{varpos}), is

\begin{equation}\label{}
  R_{true}^2 = R_\phi^2 + R_D^2 = r_T^2.
\end{equation}
Thus, while $R_\phi$ decreases with increasing transverse mass $M_T$,
$R_{true}$ does not depend on the transverse mass. This explains why the
measured HBT radii may be very small at large particle (transverse) mass, while
the \textit{actual} emission region is mass independent.

\section{Example II}
 As an instructive, more complicated, example of the phase we now consider

\begin{equation}\label{}
  \phi({\bf p}) = \frac{4}{3}\sum_i a_i^{-3} p_i^3.
\end{equation}
where $i=x,y$ (we consider only  transverse dimensions).
In order to keep the discussion simple we will
 start with the Gaussian density matrix

\begin{equation}\label{gaus}
  \rho(\textbf{p}_1,\textbf{p}_2) =\frac1{2\pi \Delta^2} e^{-\frac{\textbf{K}^2}{2\Delta^2}}e^{-\frac{1}{2}R^2\textbf{q}^2},
\end{equation}
With this choice we obtain

\begin{equation}\label{rhalai}
\rho_\phi(\textbf{p}_1,\textbf{p}_2) =
\rho(\textbf{p}_1,\textbf{p}_2)e^{i[q_x(4K_x^2+q_x^2/3)/a_x^3+q_y(4K_y^2+q_y^2/3)/a_y^3]}
\end{equation}
where we have used the identity $p_1^2 + p_1p_2 + p_2^2 = 3K^2 +
\frac{1}{4}q^2$ valid for both the $x$ and the $y$ components. Using the
formulae of the previous section we find from (\ref{gaus}) and
(\ref{rhalai}) the average shift of the {\bf X} distribution:
\begin{equation}
<X_j>(K_j)=-4K_j^2/a_j^3    \label{defom}
\end{equation}
where $j = x,y$.

From relation (\ref{rhalai}) we obtain

\begin{equation}\label{}
  S_\phi(\textbf{K},\textbf{X}) =
\int \frac{d^2\textbf{q}}{4\pi^2} e^{i\textbf{qX}}\rho_\phi(\textbf{p}_1,\textbf{p}_2) =
   s_{x\phi}(K_x,x)s_{y\phi}(K_y,y),
\end{equation}

with
\begin{equation}\label{}
  s_{j\phi}(K_j,X_j) =\frac{
e^{-\frac{K_j^2}{2\Delta^2}}}{\Delta\sqrt{2\pi}}
  \int_{-\infty}^{+\infty} \frac{dq_j}{2\pi}\exp\left[-\frac{1}{2}R^2q_j^2 + iq_jX_j + i
  \frac{1}{3a_j^3}(12K_j^2q_j + q_j^3)\right].
\end{equation}
The integral can be evaluated in terms of the Airy function.
To simplify notation, from now on we drop the subscript $j$.

Introducing a new variable $z$ by

\begin{equation}\label{}
  q = a (z - \frac{i}{2}\omega^2)
\end{equation}
where
\begin{equation}
 \omega=aR    \label{defom}
\end{equation}

we obtain

\begin{equation}\label{}
s_{\phi}({K},{X}) =\frac{ae^{-\frac{K^2}{2\Delta^2}}}{\Delta\sqrt{2\pi}}
\int_{-\infty}^{+\infty} \exp\left[\frac{i}{3}z^3
+ iAz + B\right]\frac{dz}{2\pi},
\end{equation}
where

\begin{eqnarray}\label{}
  A &=& a\left[X-<X>(K)\right]+\omega^4/4\equiv a\hat{X}+\omega^4/4,\\
  B &=&  \frac{\omega^2}{2}\left[a\hat{X}+\omega^4/6\right]
\end{eqnarray}

The integral can be expressed in terms of the Airy function (cf. \cite{ABS} for
an equivalent formula)

\begin{equation}\label{}
  Ai(u) = \frac{1}{2\pi}\int_{-\infty}^{+\infty} e^{\frac{i}{3}t^3+iut}dt
\end{equation}
The result is

\begin{equation}\label{}
  s_{\phi}({K},{X})=\frac{a}{\Delta\sqrt{2\pi}}
e^{\frac{-K^2}{2\Delta^2}}e^{B}Ai(A)
\end{equation}

One sees from this formula that, for fixed $K$, the variation of
$\hat{X}$, i.e. spread of $X$ around the average $\langle
X\rangle(K)$ is determined by two parameters: $a$ and $\omega$.
From the known asymptotic expansion of $Ai(z)$ one can deduce that
in the limit $a \rightarrow \infty$ one recovers the original
Gaussian distribution. In the limit $a \rightarrow 0$ the
distribution becomes singular. Formally one obtains
\begin{equation}
  s_{\phi}({K},{X})\; \rightarrow\; \frac{ae^{\frac{-K^2}{2\Delta^2}}}
{\Delta\sqrt{2\pi}}Ai(a\hat {X}).
\end{equation}
 and the averages of ${ \hat{X}}$ and  ${ \hat{X}}^2$ do not
exist.

\begin{figure}[htb]
\centerline{%
\epsfig{file=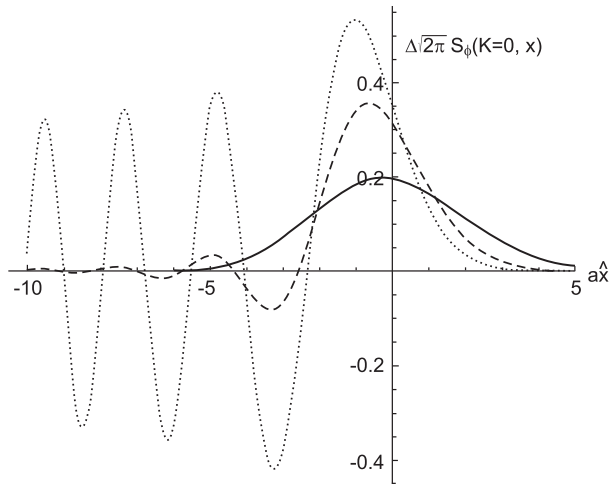,width=8cm}}
\caption{Emission function $s_{\phi}({K},{X})$ versus $a\hat{X}$. Dotted line:
$\omega=0$, dashed line: $\omega=1$, full line: $\omega=2$.}
\end{figure}

The Airy  function is negative for some values of $A$, showing  explicitly  that
the function $ s_{\phi}({K},{X})$ cannot be literally interpreted  as a distribution
function. This is illustrated in Fig. 1, where $ s_{\phi}({K},{X})$ is
plotted versus $a\hat{X}$ for $K=0$ and $\omega =0,1,2$. For $\omega=0$
it is an Airy function. With increasing $\omega$ the maximum moves
towards $\hat{X}=0$, the oscillations get dumped and already at $\omega
=2$ the curve is almost a Gaussian.

 The formulae from Section 3 yield

\begin{equation}\label{}
  \langle x_j \rangle = -\frac{4\Delta^2}{a_j^3};\qquad \sigma^2(x_j) = R^2 + 2 \langle x_j
  \rangle^2
\end{equation}
The net result is a shift of the interaction region and an increase of its
size.

\section{Conclusions}
For models where particle production is uncorrelated except for the
Bose-Einstein correlations, we have identified a group of transformations of
the phase of the single particle density matrix which leave all the the single-
and multiparticle momentum distributions invariant. The effect of the resulting
uncertainty of the size and shape of the interaction region as deduced from the
measured momentum distributions is systematically discussed. Explicit formulae
taking into account this uncertainty are derived for the moments of the
$\textbf{X}$ distribution. It is shown that the phase ambiguity can have a
dramatic effect on the parameters characterizing the interaction region. Two
explicit examples are discussed in detail.

\end{document}